\documentclass[twocolumn,showpacs,prb,amsmath,amssymb]{revtex4}

\usepackage{graphicx}
\usepackage{dcolumn}
\usepackage{bm}

\begin{document}
\title{
Critical microwave-conductivity fluctuations across the phase diagram of 
superconducting La$_{2-x}$Sr$_x$CuO$_4$ thin films 
}
\author{
Haruhisa Kitano$^1$, Takeyoshi Ohashi$^1$, Atsutaka Maeda$^1$, 
Ichiro Tsukada$^2$ 
}
\address{
$^1$Department of Basic Science, University of Tokyo, 3-8-1, Komaba, Meguro-ku, Tokyo 153-8902, Japan \\
$^2$Central Research Institute of Electrical Power Industry, 2-11-1, Iwadokita, Komae, Tokyo 201-8511, Japan
}
\date{May 30, 2005}
\begin{abstract}
We report a systematic study of the dynamic microwave conductivity 
near $T_c$ for La$_{2-x}$Sr$_x$CuO$_4$ (LSCO) thin films with 
$x$=0.07 to 0.16. 
The strong frequency dependence of the phase stiffness together with 
scaling analysis of the ac fluctuating conductivity of superconductivity 
provide direct evidence for the 2D-$XY$ behavior of nearly 
decoupled CuO$_2$ planes in underdoped LSCO ($x$=0.07 and 0.12). 
On the other hand, the critical exponents for slightly overdoped LSCO 
($x$=0.16) were found to agree with those for the relaxational 3D-$XY$ model, 
indicating that the universality class in LSCO is changed by hole doping. 
The implication of these results for the phase diagram of high-$T_c$ cuprates 
is discussed. 
\end{abstract}

\maketitle

One of the hallmarks of the high-$T_c$ cuprates is 
the large thermal fluctuation of the superconducting 
order enhanced by the short coherence length and 
the quasi two-dimensionality, which enables the exploration of 
the fluctuation-dominated critical regime very close to $T_c$ \cite{FFH91}. 
Although numerous measurements, such as the ac conductivity \cite{Kamal94etc}, 
the dc magnetization \cite{Li}, the specific heat \cite{specific_heat}, 
and the $I$-$V$ curves \cite{Strachan01}, 
have been performed to investigate the critical fluctuation, 
there has been no consensus among the results. 
This is surprising because the critical phenomena were considered to be 
universal, 
independent of the microscopic details 
\cite{ChakinTextbook}. 
However, if many assumptions were made implicitly to determine the universality class of the phase transition from the data, 
the obtained results are not convincing unless the validity of such assumptions is confirmed. 
Thus, one should develop a more reliable method 
which does not require any extra assumptions. 
To our knowledge, the most successful method is 
dynamic scaling analysis of the ac complex conductivity, 
$\sigma(\omega)=\sigma_1(\omega)-i\sigma_2(\omega)$, 
as will be discussed later. 

Another fascinating hallmark of the high-$T_c$ cuprates is 
that the physical properties change with hole doping. 
In particular, an understanding of the phase diagram as 
a plot of $T_c$ versus the hole concentration 
has been a central issue in the physics of high-$T_c$ cuprates. 
Interestingly, some recent models, 
which start from the quantum criticality for competing orders underlying the phase diagram of high-$T_c$ cuprates, provide another possible 
explanation for the critical fluctuations in high-$T_c$ cuprates. 
That is, the critical fluctuations change with hole doping, because of 
the existence of a quantum critical point (QCP) \cite{Sachedev00}. 
These models suggest that the critical dynamics should be investigated 
as a function of hole doping. 
Thus, it can be expected that such a systematic study across the phase diagram 
will not only resolve the disagreements among earlier studies 
but will also provide important information to understand 
the phase diagram of high-$T_c$ cuprates. 

In this paper, we report a systematic study of $\sigma(\omega)$ 
as a function of the swept-frequency (0.1~GHz to 12~GHz) for high-quality 
La$_{2-x}$Sr$_x$CuO$_4$ (LSCO) thin films with a wide range of 
hole concentrations ($x$=0.07 to 0.16). 
For underdoped (UD) LSCO, we show clear evidence for 
the 2D-$XY$ critical fluctuation of nearly decoupled CuO$_2$ planes. 
With increasing hole doping, a dimensional crossover from 2D-$XY$ behavior to 
3D-$XY$ behavior was observed near $x$=0.16, 
implying that there are at least two universality classes 
in the phase diagram of LSCO. 
\begin{table}[b]
 \caption{Values of $t$, $\rho_{\rm dc}$, and $T_c$ for the measured films. 
 $\rho_{\rm dc}$ is the value at $T$=50~K.
 As for three kinds of $T_c$, see the text for definition. }
 \label{tbl1}
 \begin{ruledtabular}
  \begin{tabular}{cccccc}
   $x$ & $t$~(nm) & $\rho_{\rm dc}$~(m$\Omega$cm) & $T_c^{\rm scale}$~(K) & $T_c^0$~(K) & $T_{\rm MF}$~(K) \\
   \hline
   0.07 & 460 & 0.77 & 19.0 & 20.83 & $\sim$32 \\
   0.12 & 230 & 0.28 & 33.65 & 36.08 & $\sim$38 \\
   0.14 & 270 & 0.14 & 38.92 & 39.29 & $\sim$40 \\
   0.16 & 140 & 0.12 & 35.5 & 35.82 & --- \\
  \end{tabular}
 \end{ruledtabular}
\end{table}

Epitaxial LSCO thin films with $x$=0.07, 0.12 (underdoped), 0.14 
(nearly optimally doped), and 0.16 (overdoped) were grown on 
LaSrAlO$_4$ (001) substrates by a pulsed laser deposition technique 
using pure ozone \cite{Tsukada04}. 
All the films are highly $c$-axis oriented with a sufficiently narrow 
rocking curve of the 002 reflection (typically 0.2${}^{\circ}$). 
As shown in Table I, the value of the in-plane dc resistivity, 
$\rho_{\rm dc}$, was found to agree with the reported best value for 
LSCO thin films \cite{Sato97} within a factor of 2, confirming that 
the films used in this study are of sufficiently high quality 
to investigate the critical dynamics near $T_c$. 

There are three reasons for using LSCO films on LaSrAlO$_4$ (LSAO) 
substrates in this study: 
(1) LSCO is an ideal system with a simple layered structure, 
where the hole concentration can be widely controlled. 
(2) The compressive epitaxial strain gives rise to a moderate increase 
of $T_c$ 
\cite{Sato97}. 
(3) The tetragonal symmetry of LSAO substrate supports the fabrication of 
CuO$_2$ planes with ideal flat square lattices, free from disorders 
due to corrugations and twin boundaries \cite{Tsukada04}. 
%
%

Both the real and imaginary parts of $\sigma(\omega)$ were 
obtained from the complex reflection coefficient, 
$S_{11}(\omega)$, using a non-resonant broadband technique 
\cite{Kitano04}. When the film thickness, $t$, is sufficiently less 
than the skin depth, $\delta$, one can write $\sigma(\omega)$, as follows, 
\begin{equation}
\sigma(\omega)=\frac{1}{tZ_0}\frac{1-S_{\rm 11}(\omega)}{1+S_{\rm 11}(\omega)},
\label{eq1}
\end{equation}
where $Z_0$=377 $\Omega$ is the impedance of free space. 
In practice, before applying Eq.~(\ref{eq1}), systematic errors involved in 
$S_{11}(\omega)$ were carefully removed by calibration measurements 
using known standards \cite{Kitano04}. 
\begin{figure}[b]
  \begin{center}
    \includegraphics[width=8.5cm]{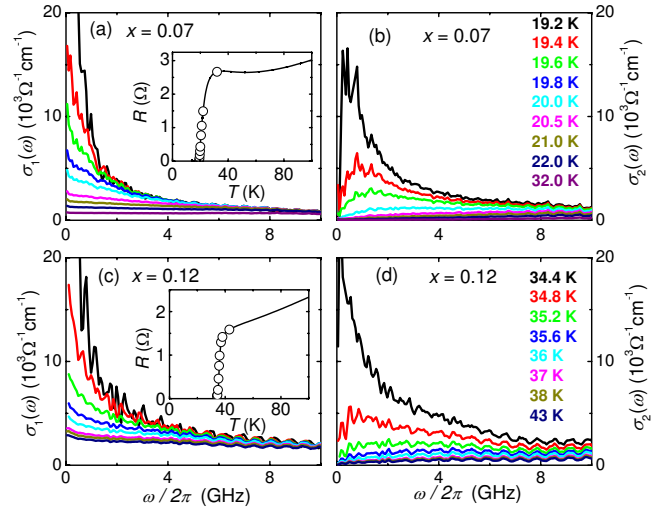}
     \caption{The frequency dependence of (a) $\sigma_1(\omega)$ for $x$=0.07, 
     (b) $\sigma_2(\omega)$ for $x$=0.07, (c) $\sigma_1(\omega)$ for $x$=0.12, 
     (d) $\sigma_2(\omega)$ for $x$=0.12, respectively. 
     Insets: the temperature dependence of the dc resistance, $R$, of the same 
     sample. Open circles near $T_c$ show $R$ at the temperatures presented 
     in the main panel. 
     }
    \label{fig1}
  \end{center}
\end{figure}
\begin{figure}[b]
  \begin{center}
    \includegraphics[width=7.5cm]{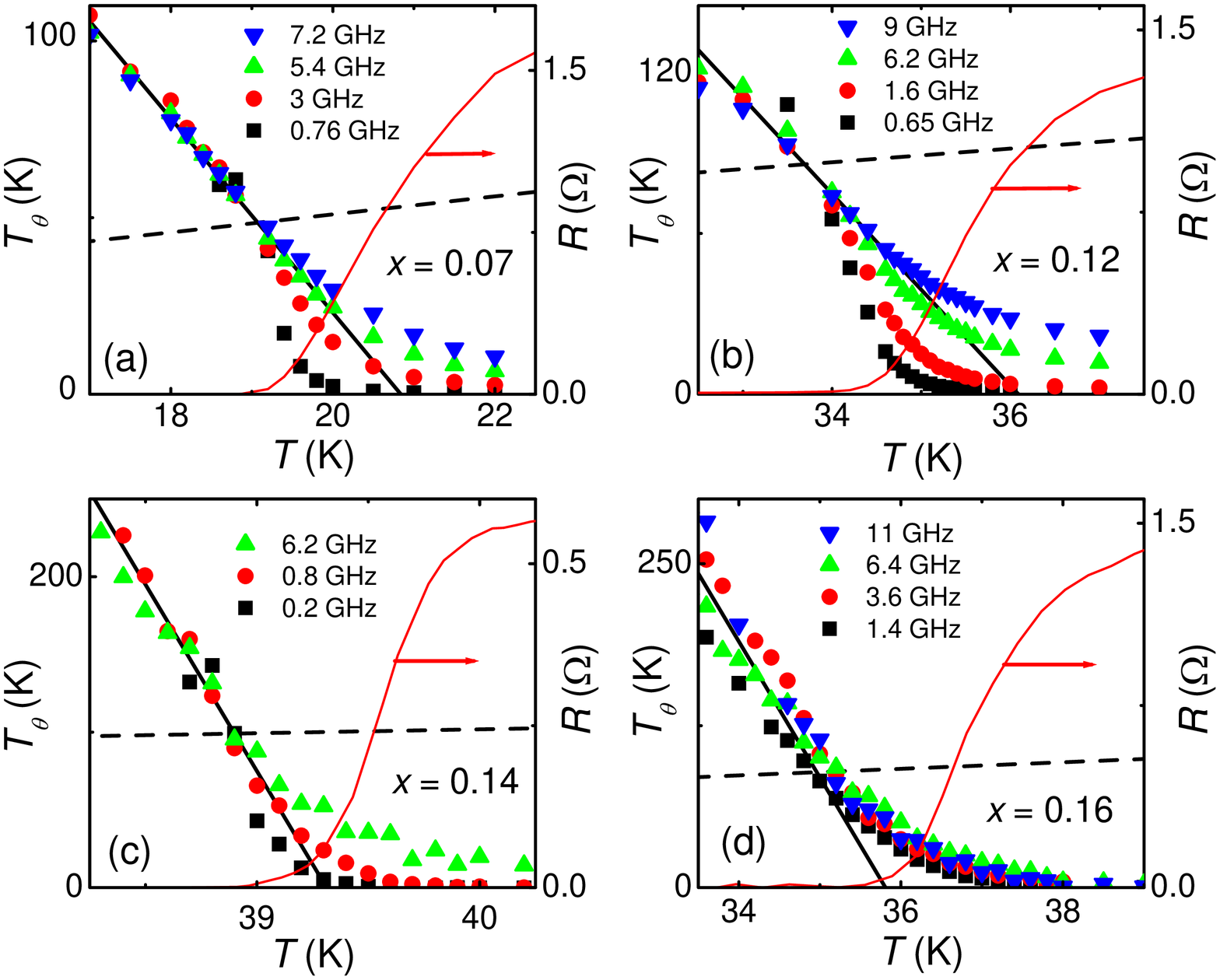}
     \caption{
     The phase-stiffness temperature for the film with (a) $x$=0.07, 
     (b) $x$=0.12, (c) $x$=0.14, and (d) $x$=0.16. 
     The dashed straight line represents $(8/\pi)T$. 
     The solid straight line gives the bare phase stiffness in the $XY$ model, 
     while it gives the mean-field superfluid density in the GL theory. 
     In each panel, the dc resistance was given by red solid curves. 
     }
    \label{fig2}
  \end{center}
\end{figure}

In order to obtain reliable $\sigma(\omega)$ data, 
the effect of the dielectric substrate needs to be considered. 
Note that the choice of $t$ according to values of $\rho_{\rm dc}$ is 
important, because a resonance peak attributed to 
the effect of substrate can be induced for a very small value of $t$ 
\cite{Silva96}. 
In fact, a sharp resonance peak has been observed near 8~GHz for 
the LSCO film ($x$=0.07) above 100~K \cite{Kitano04}. 
However, in the data presented in Fig.~1, 
no peak was observed over the whole frequency range 
measured at lower temperatures near $T_c$. 
Thus, we can conclude that the effect of substrate was negligible 
near $T_c$. 

Moreover, in the vicinity of $T_c$, we found that 
$\sigma(\omega)$ was seriously affected by 
a small unexpected difference (typically, 0.2$^\circ$-0.3$^\circ$ at 1~GHz) 
between the phase of $S_{11}$ for a load standard, $\angle S_{11}^{\rm load}$, 
and that for a short standard, $\angle S_{11}^{\rm short}$. 
This difficulty was resolved by using the measured $S_{11}(\omega)$ 
of the sample at a temperature far above $T_c$ as the load standard, 
assuming that 
$\angle S_{11}^{\rm short}=\angle S_{11}^{\rm load}$ at this temperature, 
and that both were $T$-independent in the vicinity of $T_c$. 
The validity of this procedure was also confirmed by the same measurements 
for NbN thin films as a reference \cite{OhashiNbN}. 

Figure 1 shows the frequency dependence of $\sigma(\omega)$, obtained by 
the above procedures, for underdoped LSCO ($x$=0.07 and 0.12) 
at several temperatures above $T_c$. 
It is evident that both $\sigma_1(\omega)$ and 
$\sigma_2(\omega)$ diverge rapidly with decreasing temperature 
in the low frequency limit, suggesting that the excess conductivity is 
due to the superconducting fluctuations. 
A similar divergence in $\sigma(\omega)$ was also observed 
in the vicinity of $T_c$ for the other LSCO films. 

It has been argued \cite{EK95} that 
strong phase fluctuations are important for 
the determination of $T_c$ in the UD region, 
because the phase-stiffness energy, 
$k_BT_\theta\equiv(\hbar/e^2)\hbar\omega\sigma_2d_s$, 
is suppressed largely by the small superfluid density, 
where $d_s$ is the effective thickness of a superfluid. 
Figure 2 shows the phase-stiffness temperature $T_\theta(T)$ estimated from 
$\sigma_2(\omega,T)$ at several frequencies. 
Surprisingly, we found that $T_\theta(T)$ for the UD samples ($x$=0.07 
and 0.12) started to show frequency dependence above a certain temperature, 
$T_k$, while $T_\theta(T)$ was almost $\omega$-independent below $T_k$. 
In addition, the resistive $T_c$ ($T_c^R$), where $R$ became zero, 
was also close to $T_k$ rather than 
another critical temperature, $T_c^0$, where the bare phase stiffness 
in the $XY$ model would go to zero. 
When we assumed $d_s$ to be approximately $t/2$, 
we saw that a dashed straight line with a slope of $8/\pi$ crossed 
$T_\theta(T)$ at $T_k$, indicating that $T_k$ agrees with 
the Berezinskii-Kosterlitz-Thouless (BKT) transition temperature, 
$T_{\rm BKT}$, in the 2D-$XY$ model \cite{KT,KT2}. 

As shown in Figs.~2(c) and 2(d), this behavior was observed even for 
$x$=0.14, while it disappeared almost completely for $x$=0.16. 
$T_c^R$ for $x$=0.16 was close to $T_c^0$ rather than 
$T_{\rm BKT}$, suggesting that the mean field critical temperature 
$T_{\rm MF}$ is close to $T_c^0$. 
Thus, our results were qualitatively similar to the prediction of 
Emery and Kivelson \cite{EK95} in the sense 
that $T_c$ is bounded by $T_{\rm BKT}$ (UD region) or 
$T_{\rm MF}$ (overdoped region), 
although the estimation of $T_{\rm MF}$ in the UD region was considerably 
different, as will be discussed later. 

In the dynamic scaling analysis of 
the fluctuating complex conductivity, $\sigma_{\rm fl}(\omega)$, 
which was pioneered by Booth {\it et al.} \cite{Kamal94etc}, 
both the magnitude, $|\sigma_{\rm fl}|/\sigma_0$, and the phase, 
$\phi_\sigma(\equiv$$\tan^{-1}[\sigma_2^{\rm fl}/\sigma_1^{\rm fl}])$, 
of $\sigma_{\rm fl}(\omega)$ are used as scaled quantities. 
The advantage of this method is that the 
data collapse to a single curve as a function of a reduced frequency, 
$\omega/\omega_0$, can be achieved without assuming any relationship between 
two scaling parameters, $\sigma_0$ and $\omega_0$, in contrast to 
other scaling analyses \cite{Li,specific_heat,Strachan01}. 
Note that $\sigma_0$ and $\omega_0$ are obtained independently 
in our analysis. 
Thus, we can begin by checking the following hypothesis of 
the dynamic scaling theory 
\cite{FFH91}, 
\begin{equation}
\sigma_{\rm fl}(\omega)\approx \xi^{z+2-d}S(\omega\xi^z),
\label{eq2}
\end{equation}
where $S(x)$ is a complex universal scaling function, $\xi$ is a 
correlation length which diverges at $T_c$, 
$z$ is a dynamic critical exponent, 
and $d$ is an effective spatial dimension. 
Figure 3 shows that both $|\sigma_{\rm fl}|/\sigma_0$ and $\phi_\sigma$ 
were scaled successfully over a wide range of frequencies for all the films, 
confirming that the critical dynamics suggested by Eq.~(\ref{eq2}) 
were indeed observed. 
The comparison of the experimentally obtained scaling functions with 
the Gaussian forms calculated by Schmidt \cite{Schmidt} suggested that 
the 2D Gaussian-like behavior in the UD region 
changed into the 3D Gaussian-like behavior near $x$=0.16. 

Equation~(\ref{eq2}) suggests that both $\sigma_0$ and $\omega_0$ behave as 
functions of $\xi$ in a critical region, where 
$\sigma_0\propto\xi^{z+2-d}$ and $\omega_0\propto\xi^{-z}$. 
According to the BKT theory \cite{KT,KT2}, 
$\xi_{\rm KT}$ diverges with $\exp[b/\sqrt{\epsilon}]$ 
in the critical region, $T_c<T\ll T_c^0$, 
where $\epsilon\equiv T/T_c^{\rm scale}-1$ and $b$ is a numerical constant. 
Note that $\sigma_0=1/\omega_0\propto\xi^2$ for the relaxational 2D system 
with $z$=2 \cite{FFH91}. 
In fact, we confirmed that both $\sigma_0$ and $\omega_0$ for $x$=0.07 
showed the same exponential singularity as $\xi_{\rm KT}^2$ with $b$=0.215 
in the range of $\epsilon$ from 0.01 to 0.1, as shown in Figs.~4(a) and 4(b). 
With increasing $x$ up to 0.14, we found that the range of $\epsilon$ where 
$\sigma_0$ (or $\omega_0$) agreed with $\xi_{\rm KT}^2$ became 
narrower and shifted to lower temperatures. 
Such behaviors were also consistent with the BKT theory \cite{KT2}, 
since $\sigma_0$ (or $\omega_0$) is rather dominated by 
free vortices than $\xi_{\rm KT}^{2}$ at higher temperatures, 
due to a screening effect by thermally activated free vortices. 
\begin{figure}[b]
  \begin{center}
    \includegraphics[width=8.5cm]{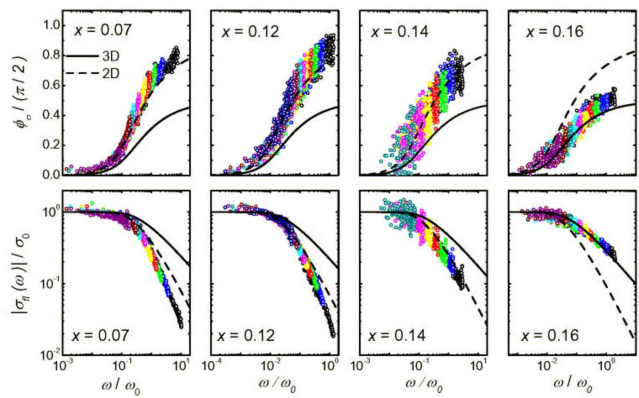}
     \caption{
     Scaled curves of $\phi_\sigma$ (upper panels) and 
     $|\sigma_{\rm fl}|$ (lower panels) for the measured LSCO film. 
     $\sigma_{\rm fl}(\omega)$ was obtained by subtracting the normal-state 
     conductivity at 32~K ($x$=0.07), 38~K ($x$=0.12) and 
     40~K ($x$=0.14, 0.16), respectively. 
     Data for $x$=0.07 span the frequency range 0.2$\sim$7~GHz at reduced 
     temperatures $\epsilon$=0.01$\sim$0.5, while those for $x$=0.16 span 
     0.2$\sim$2~GHz at reduced temperatures, 
     $\epsilon$=0.002$\sim$0.08. 
     The solid (dashed) lines are the 3D (2D) Gaussian scaling functions. 
     }
    \label{fig3}
  \end{center}
\end{figure}
\begin{figure}[t]
  \begin{center}
    \includegraphics[width=8.0cm]{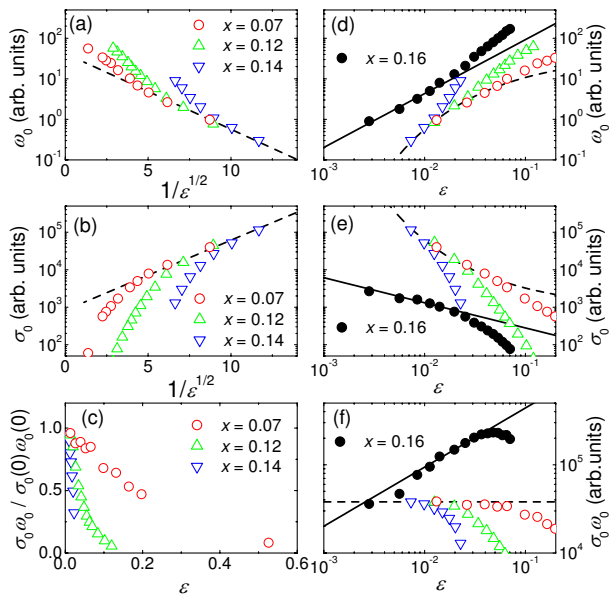}
     \caption{
     (a) $\omega_0$ and (b) $\sigma_0$ as a function of $1/\sqrt{\epsilon}$ 
     for $x$=0.07 to 0.14, where $\epsilon=T/T_c^{\rm scale}-1$. 
     The dashed line is (a) $\xi_{\rm KT}^{-2}$ and (b) $\xi_{\rm KT}^{2}$. 
     (c) $\sigma_0\omega_0$ vs $\epsilon$ for $x$=0.07 to 0.14.
     (d) $\omega_0$, (e) $\sigma_0$ and (f) $\sigma_0\omega_0$ 
     as a function of $\epsilon$ for $x$=0.07 to 0.16. 
     The solid line is (d) $\epsilon^{1.33}$, (e) $\epsilon^{-0.67}$ 
     and (f) $\epsilon^{0.67}$. 
     The dashed line is (d) $\xi_{\rm KT}^{-2}$, (e) $\xi_{\rm KT}^{2}$ 
     and (f) the behavior expected in $d$=2. 
     Note that the absolute values of the plots in all the panels 
     except for (c) are unimportant, 
     since they include arbitrary proportional coefficients. 
     }
    \label{fig4}
  \end{center}
\end{figure}

In the BKT theory, $\sigma_0\omega_0$ gives $T_\theta$ in 
the high-frequency limit, $T_\theta^0$, 
which is sensitive to the surviving bound pairs of vortices above 
$T_{\rm BKT}$ \cite{KT2}. 
As shown in Fig.~4(c), we found that 
$\sigma_0\omega_0$ decreased with increasing $T$ more quickly 
at larger $x$. 
This suggests that the temperature region of the prominent phase fluctuation 
became narrower with hole doping, in contrast to 
a recent Nernst experiment\cite{Nernst} which showed a steeper increase of 
the onset temperature of the Nernst effect than $T_c$ 
with increasing hole doping. 
When we roughly estimated $T_{\rm MF}$, based on an expectation that 
$\sigma_0\omega_0\approx$0 at $T_{\rm MF}$, we found that 
$T_{\rm MF}$ was too small to cover the pseudogap region, 
as was previously reported by Corson {\it et al.} 
\cite{Kamal94etc}. 
These results strongly suggested that most of the anomalous Nernst signal 
should be attributed to other origins than the superconducting fluctuation, 
as was suggested by some recent theoretical works 
\cite{Honerkamp}. 

Moreover, the observed reduction of $T_c$ due to the phase fluctuation 
($T_{\rm BKT}/T_{\rm MF}$$\sim$ 0.7, 0.9 for $x$=0.07, 0.12, 
respectively) corresponded to the superconducting film with 
a very high sheet resistance $R_{\rm sq}$($\sim$3-5~k$\Omega$) 
\cite{Beasley79}. 
Using a relation $R_{\rm sq}$=$\rho_{\rm dc}/D_s$, we found that 
$D_s$$\sim$10 \AA. 
Thus, each of the CuO$_2$ planes seemed to be decoupled 
\cite{FFH91,Hikami80}. 
Note that the screening length for each superconducting sheet, 
$\lambda_\perp(=\lambda^2/D_s)$, where $\lambda$ is a bulk penetration depth, 
was larger than 10~mm in this case, 
which satisfied the BKT criterion that $\lambda_\perp\gg L$, 
where $L$ is the sample size. 
In addition, $d_s$ is given by a sum of the decoupled superconducting 
layers with the thickness of $D_s$. 
Thus, $d_s$ will be smaller than $t$, as we assumed in the estimation of 
$T_\theta(T)$ shown in Fig.~2. 

We found that the critical behavior for $x$=0.16 was very different from 
that for $x$=0.07-0.14, as shown in Figs.~4(d) to 4(f). 
In particular, the dimensionality of $x$= 0.16 was found to be three, 
in contrast to the 2D-$XY$ behavior of $x$= 0.07-0.14, 
as shown by the plots of 
$\sigma_0\omega_0(\propto\epsilon^{\nu(d-2)}$) in Fig.~4(f), 
where $\nu$ is a static critical exponent. 
We also confirmed that the critical exponents for $x$=0.16 agreed very well 
with those for the relaxational 3D-$XY$ model with 
$\nu$$\approx$0.67, $z$$\approx$2, and $d$=3 \cite{Wickham00}. 
These results clearly suggest that the universality class in 
the pahse diagram of LSCO changes from 2D-$XY$ 
to 3D-$XY$ with hole doping. 

What is the origin of the dimensional crossover from 2D-$XY$ to 3D-$XY$ ? 
One of candidates is the increase of the interlayer coupling 
(probably Josephson coupling) with hole doping \cite{FFH91,Hikami80}. 
However, this effect seems to appear more gradually with hole doping, 
in contrast to the different behavior between $x$=0.14 and 0.16 
shown in Figs.~3 and 4(f). 
Another candidate is the effect of the quantum critical fluctuation near QCP. 
In this case, the 2D-3D crossover can be regarded as 
the classical-to-quantum crossover near QCP \cite{Sachedev00}. 
A detailed study for further overdoped LSCO is needed to settle this issue. 

Finally, we emphasize that all the above results rule out the possibility 
of the distribution of $T_c$ due to disorders in the sample. 
If there is a broad distribution of $T_c$ in the sample, 
$\sigma_2$ will be greatly suppressed, leading to a smaller $T_\theta$, and 
the dynamic scaling of $|\sigma|$ and $\phi_\sigma$ will fail below a 
certain temperature, suggesting that $T_c^{\rm scale}$ is higher than $T_c^R$. 
In fact, these features have been observed for NbN films with 
a non-negligible distribution of $T_c$ \cite{OhashiNbN}, 
while all of the four LSCO films used for this study did not show them, 
as shown in Figs.~2 to 4. 

In conclusion, a systematic study of the critical dynamics of 
$\sigma_{\rm fl}(\omega)$ for LSCO thin films with $x$=0.07 to 0.16 
has been performed for the first time. 
All the results clearly provide evidence for the BKT transition 
in the nearly decoupled CuO$_2$ layers of underdoped LSCO. 
With increasing hole doping, the 2D-$XY$ behavior in the UD region 
changes into the 3D-$XY$ behavior near $x$=0.16, 
indicating that the universality class in the phase diagram of LSCO 
is changed by hole doping. 

We thank Y. Kato, H. Fukuyama and W. N. Hardy 
for fruitful discussions and comments, 
S. Anlage and A. Schwartz 
for technical advice at the early stages of this study, 
D. G. Steel and K. Gomez for useful comments on the manuscript. 
This work was partly supported by the Grant-in-Aid for Scientific Research 
(13750005, 14340101 and 15760003) 
from the Ministry of Education, Science, Sports and Culture of Japan. 


%
%

\end{document}